\documentclass{iau} 
\usepackage{graphicx}

\title[MULTIFRACTAL SIGNATURES OF GRAVITATIONAL WAVES] 
{MULTIFRACTAL SIGNATURES OF GRAVITATIONAL WAVES DETECTED BY LIGO}

\author[Daniel B. de Freitas et al.]   
{Daniel B. de Freitas$^1$
 \and Mackson M. F. Nepomuceno$^2$
\and J. R. De Medeiros$^3$}

\affiliation{$^1$Departamento de F\'{\i}sica, Universidade Federal do Cear\'a, \\ Caixa Postal 6030, Campus do Pici,  
	60455-900 Fortaleza, Cear\'a, Brazil \\ email: {\tt danielbrito@fisica.ufc.br} \\[\affilskip]
$^2$Departamento de Ci\^{e}ncia e Tecnologia, Universidade Federal Rural do Rio Grande do Norte-UFERSA,
\\ Campus Cara\'ubas, Rio Grande do Norte, Brazil
$^3$ Departamento de F\'{\i}sica Te\'{o}rica e Experimental, \\ Universidade Federal do Rio Grande do Norte-UFRN, 
Rio Grande do Norte, Brazil}

\pubyear{2020}
\volume{346}  
\jname{High-mass X-ray binaries: illuminating the passage from massive binaries to merging compact objects}
\editors{L. Oskinova, ed.}
\begin{document}

\maketitle

\begin{abstract}
We analyze the data from the 6 gravitational waves signals detected by LIGO through the lens of multifractal formalism using the MFDMA method, as well as shuffled and surrogate procedures. We identified two regimes of multifractality in the strain measure of the time series by examining long memory and the presence of nonlinearities. The moment used to divide the series into two parts separates these two regimes and can be interpreted as the moment of collision between the black holes. An empirical relationship between the variation in left side diversity and the chirp mass of each event was also determined.
\keywords{gravitational waves --- methods: statistical}

\end{abstract}

\firstsection 
\section{Introduction}

Since the first detection, five more signals have been confirmed as GWs: GW151226 (\cite{GW2andLVT}), GW170104 (\cite{GW3}), GW170608 (\cite{GW170608}), GW170814 (\cite{GW170814}) and GW170817 (\cite{GW170817}) (the only one coming from a system of coalescing neutron stars), and one signal remains as a suspected GW (LVT151012 \cite{GW2andLVT}). The GW data used here are within the range of 32 seconds around the event and have a measurement frequency of 4096Hz. We will assume that these GWs, denoted by $y(t)$, are linear combinations of a deterministic signal, $d(t)$, and background noise, $n(t)$. In this context, the present analysis deals with observations that are collected over evenly spaced and discrete time intervals. In this Letter, we reports an analysis of a search for traces of multifractality in GW150914, a fact that may have strong consequences for our understanding of different characteristics of GW. A general discussion of all the GW signals detected to date (with the exception of GW170817) will be also presented. The signal of GW170817 (from coalescence of binary neutron stars) was removed from the sample since it differs in number of data from the signals produced by coalescence of black holes.

\section{Multifractal analysis} \label{sec:style}

In monofractal series, one exponent (the Hurst exponent \cite{Hurst}) is sufficient to characterize the behavior of the series at various scales. $H$ values of $0<H<0.5$ and $0.5<H<1$ indicate persistence and anti-persistence, respectively, while $H=0.5$ indicates that the time series is uncorrelated. In multifractal time series, a range of values for this exponent is calculated. Thus, multifractal analysis consists of studying the scaling behavior in the time series $y(t)$. First, in accordance with the MultiFractal Detrending Moving Average (MFDMA) procedure, we calculated the mean-square function $F^2_\nu(n)$ for a $\nu$ segment of size {\it n}:

\begin{equation}\label{fluctuMS}
F^2_\nu(n)=\frac{1}{n}\sum_{i=1}^{n}[e_\nu(i)]^2,
\end{equation}
where $e_\nu(i)=y(i)-\tilde{y}(i)$ is the residual series in the segment $\nu$ and $\tilde{y}(i)$ is the moving average function. However, some authors have shown that semi-sinusoidal and power-law trends in multifractal approaches, including Multifractal Detrended Fluctuation Analysis (MFDFA) and MFDMA, are not efficiently removed (\cite{egh}). In our study, we did not encounter this problem. We then calculated the $q_{th}$ order overall fluctuation function $F_q(n)$, which is given by

\begin{equation}\label{fluctu}
F_q(n)=\left\lbrace \frac{1}{N_n} \sum_{\nu=1}^{N_n} F^q_\nu(n) \right\rbrace^{1/q} \textrm{for}\; q\neq0
\end{equation}
and, for $q=0$,
\begin{equation}\label{fluctuzero}
\ln\left[F_{0}(n)\right]=\frac{1}{N_{n}}\sum^{N_{n}}_{\nu=1}\ln [F_{\nu}(n)],
\end{equation}
where $N_n$ is the number of segments non-overlaping. For larger values of $n$, the fluctuation function follows a power-law given by

\begin{equation} \label{Fqxn}
F_q(n) \sim n^{h(q)}.
\end{equation}

The generalized Hurst exponent\textbf{ $h(q)$ }is related to standard multifractal analysis parameters such as the Renyi scaling exponent ($\tau$), which is given by

\begin{equation}\label{tau}
\tau(q)=qh(q)-1,
\end{equation}
\textbf{when $q = 2$, we return to using monofractal analysis, i.e., $h(2) = H$ is the Hurst exponent.}

Two other important parameters are obtained using a Legendre transform, defined as 
\begin{equation}\label{alpha}
\alpha=\frac{d\tau(q)}{dq}, \quad \alpha\in[\alpha_{min},\alpha_{max}]
\end{equation}
and
\begin{equation}\label{falpha}
f(\alpha)=q\alpha-\tau(q),
\end{equation}
which are the H\"{o}lder exponent and singularity spectrum, respectively.

One way to measure the degree of multifractality $(\Delta\alpha)$ in a series is by using the width of the multifractal singularity spectrum, which \cite{W2} and \cite{W1} defined as the difference between the maximum and minimum values of the H\"{o}lder exponent, i.e., $\Delta\alpha=\alpha_{max} - \alpha_{min}$.

\begin{figure}[!h]
	\centering
	\includegraphics[trim=100 240 70 240,clip,scale=0.9]{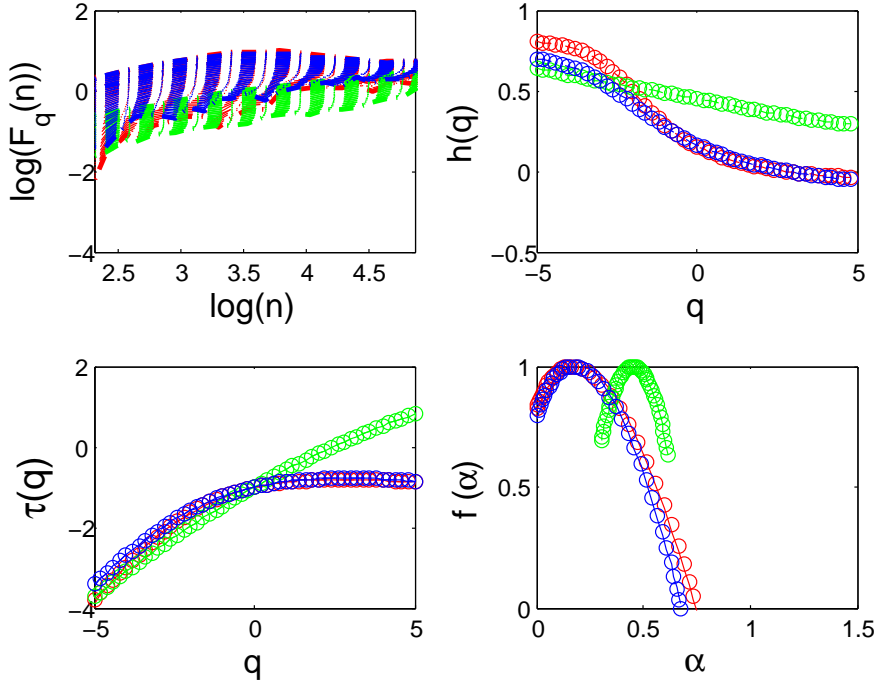}
	\caption{\label{H1data} Multifractal analysis of the full data from GW150914 (\texttt{H1data}). The top-left panel shows the fluctuation function versus the multi-scale  behavior in a log-log diagram. The original series is in red, the shuffled series is in green, and the upper and lower limits correspond to $q=5$ and $q=-5$, respectively, while the bold in the middle corresponds to $q=0$. Dependences on the $q_{th}$ moment of the generalized Hurst exponent, $h(q)$, and the multifractal scaling exponent, $\tau(q)$, are shown in the top-right and bottom-left panels, respectively. The multifractal spectrum is shown in the bottom-right panel.}
\end{figure}

\begin{figure}[!h]
	\centering
	\includegraphics[trim=90 150 90 150,clip,scale=0.9]{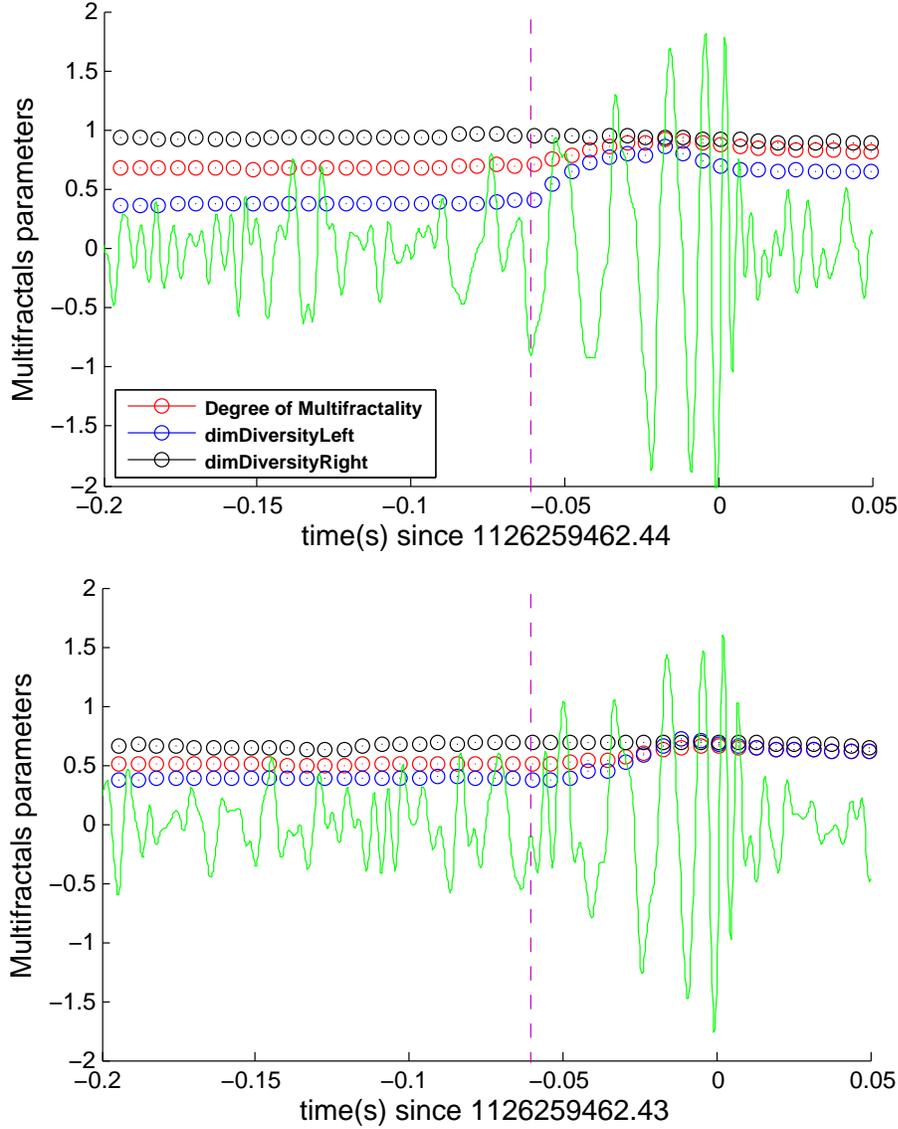}
	\caption{\label{cumulative} Point-to-point multifractal analysis for the GW150914 time series from Livingston (bottom panel) and Hanford (top panel; shifted and inverted \cite{GW}), illustrated in green. Red circles represent the degree of multifractality ($\Delta\alpha$) calculated in the time series up to that point; likewise, blue and black circles represent the left side diversity $f(\alpha)_{max}-f(\alpha)^{left}_{min}$ and right side diversity $f(\alpha)_{max}-f(\alpha)^{right}_{min}$, respectively. The vertical lines represent $t = -0.06s$, the time point at which the time series are divided.}
\end{figure}

\begin{figure}[!h]
	\centering
	\includegraphics[trim=90 220 80 220,clip,scale=0.9]{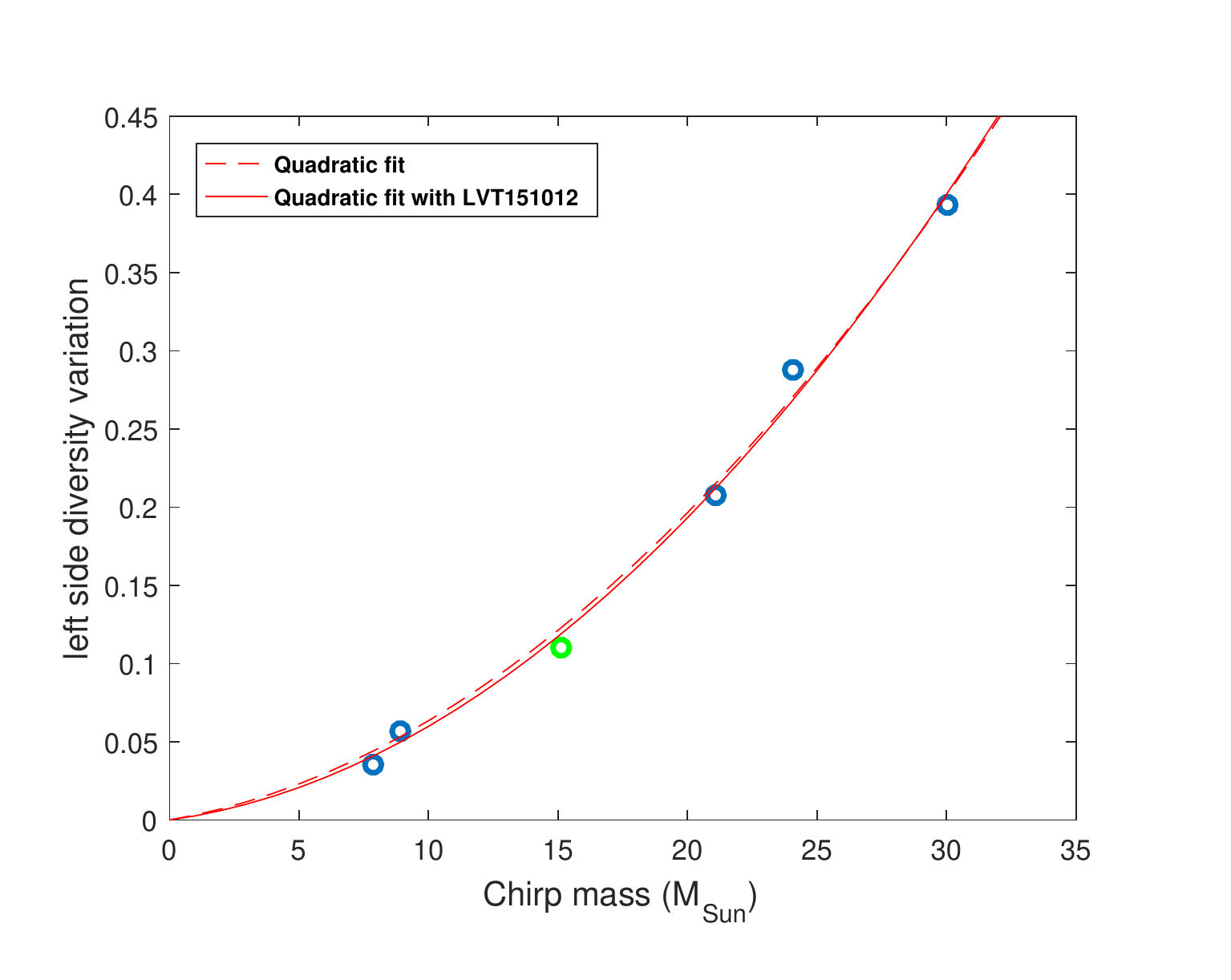}
	\caption{\label{DeltaDLMc}The correlation between left side diversity variation $(\Delta D_L)$ and chirp mass for each GW (circles in blue) and LVT151012 (circle in green). The dashed line indicates a quadratic fit adjustment without the LVT signal, and the solid line is the same fit when considering this signal.}
\end{figure}

\section{Results and discussion}

\textbf{We analyze the data from the 6 gravitational waves signals detected by LIGO identified as GW151226 , GW170104, GW170608, GW170814, GW170817 and LVT151012. All of data were extracted from LIGO. Data were analyzed using the multifractal formalism. Our aim is to study the possible sources of multifractality and to extract a set of multifractality indexes.}

To investigate the source of multifractality, we applied the shuffled method to the original series (the green curves in Figure \ref{H1data}). This method destroys the memory signature, but preserves the distribution of the data with $h(q)=0.5$, if the source of multifractality in time series only presents long-range correlations (\cite{deFreitas2}). We realized that the multifractal behavior remains but with lowered strength. Similarly, the surrogate method (the blue curves in Figure \ref{H1data}) also could not eliminate the multifractality in the original series. Already, this method destroys effects of non-linearity of the original series by randomizing the Fourier phases. These results indicate that the source of the multifractality is not only related to long-range correlations but also linked to the existence of non-linear terms that produce a heavy-tailed probability density function (PDF). The same analysis described in the previous three paragraphs was applied to the other three waves and indicated similar behavior both for the Hanford and Livingston detector data. 

To study the evolution of the parameters related to the multifractal singularity spectrum throughout the time series, we constructed Fig. \ref{cumulative}, with the original time series shown in green, for the Hanford data in the top panel and Livingston in the bottom panel. The parameter values at one point in the time series data reflect the values calculated up to that point in a 50-point data window. As seen in Fig. \ref{cumulative}, the left side diversity $(\Delta f_{L}(\alpha))$ of the multifractal singularity spectrum, defined as $1-f(\alpha)^{left}_{min}$, is shown in blue and is associated with the sensitivity of the series to small-scale fluctuations with large magnitudes. In the same Figure, the right side diversity $(\Delta f_{L}(\alpha))$ of the the multifractal singularity spectrum, denoted by $1-f(\alpha)^{right}_{min}$, is shown in black and is linked to the sensitivity to fluctuations in the series with small magnitudes \cite{W2}. Furthermore, the parameters $(\Delta f_{L}(\alpha))$ and $(\Delta f_{R}(\alpha))$ indicate either a left or right truncation of the multifractal spectrum, respectively For this analysis, the parameters were calculated for the signal in the interval between 1 second before the event (for GW150914, this time is 1126259462.44s) and 0.05s after the event. We can observe a slight increase in the left side diversity at $t=-0.06s$, which indicates the presence of a strong small-scale fluctuation. This behavior appears in the data analysis of the two advanced LIGO detectors, H1 and L1. Using this time point, we divided the original series into two parts, wherein the first is identified as \texttt{H1data1} with 3581 measurements, while the second part comprises 720 measurements and is identified as \texttt{H1data2} data.

Using the same procedure as that for the entire time series, we performed multifractal analyses on both the \texttt{H1data1} and \texttt{H1data2} data. The shuffled method has eliminated the multifractality contained in the \texttt{H1data1}, shown in the right pane, and for the shuffled series, $h(2) = 0.5429$ and $\Delta\alpha = 0.0288$. These results indicate that the multifractality present in \texttt{H1data1} is due only to long-term correlations and thus does not provide non-linear terms. These correlations can be understood as stemming from the periodic orbital motion of the black holes. As for \texttt{H1data1}, multifractality is still present for the original time series, but neither the shuffled nor surrogate methods could eliminate the multifractal behavior; i.e., the multifractal behavior is due to two possible sources, i.e., memory and non-linearity. 

These results have two consequences: first, the entire contribution of non-linearity in the analysis of the complete time series occurs in the second part of the series; second, as the periodic movement continues, even in the ringdown phase, the terms associated with long-term correlations continue to appear in the series. The enlargement of the PDF is because the amplitude of the strain grows somewhat in the second part of the series. Given that the strain amplitude is linked to the orbital velocity and mass that generated the gravitational wave, these nonlinear terms are caused by the collision of the black holes. In short, the contribution of long-term temporal correlation is due to the periodic motion of the orbiting black holes, and nonlinear terms occur due to the increase in the strain amplitude.

The difference, presented in the Figure \ref{cumulative}, between the maximum and minimum value of the left side diversity in the GW amplitude region of increase can be considered as the variation in left side diversity $(\Delta f_{L}(\alpha))$, as indicated in Figure \ref{cumulative} for GW150914. We find an empirical correlation between this parameter and the chirp masses of each signal. Figure \ref{DeltaDLMc} illustrates these parameters in blue circles for GWs and green circles for the LVT. A quadratic fit with (solid line) and without (dashed line) the LVT151012 signal is also shown in the same Figure. The overlap of these lines indicates that the analysis of the LVT signal falls within the expected behavior according to this correlation. Since we associate the variation in left side diversity with the amplitude increase in the signal, which in turn is related to chirp mass, we are led to conclude that this is an expected correlation. The detection of new GWs can serve as a good test for the correlation found here as well as a check for the chirp mass value of the detected signal.

\section{Conclusions}

The statistical approach proposed in this study highlights the scenario opened by detection of the first GWs. We summarize the main results in three points: i) characterize the fractal dynamics of the signals, identifying their multifractal sources; ii) find the moment of the beginning of merger phase in black hole coalescence system, and; iii) determine the empirical relationship between the variation in left side diversity and chirp mass as an additional way for estimating this latter parameter. The methodology applied here may serve as a standard procedure for future analyses of gravitational waves. The prospect of new gravitational wave observatories, both on the ground and in space, provides more opportunities for the field of astronomy to employ the statistical tools already widely used in other areas of knowledge.

\end{document}